\documentclass[aps,prd,nofootinbib,supercriptaddress]{revtex4}

\usepackage{amssymb,amsmath}
\usepackage{graphicx}
\usepackage{color}
\definecolor{redpantsred}{rgb}{0.8,0,0}
\definecolor{redpantsgrey}{rgb}{0.5,0.5,0.5}
\definecolor{redpantsdarkred}{rgb}{0.6,0,0}
\usepackage[
,colorlinks=true
,urlcolor=redpantsred
,anchorcolor=redpantsred
,citecolor=redpantsred
,filecolor=redpantsred
,linkcolor=redpantsred
,menucolor=redpantsred
,pagecolor=redpantsred
,linktocpage=true
,pdfproducer=redpants
,pdfusetitle
]{hyperref}
\newcommand*{\doi}[1]{\href{http://dx.doi.org/\detokenize{#1}}{{\color{redpantsdarkred} [doi]}}}
\newcommand{\beq}{\begin{equation}}
\newcommand{\eeq}{\end{equation}}
\newcommand{\bea}{\begin{eqnarray}}
\newcommand{\eea}{\end{eqnarray}}
\newcommand{\vc}[1]{{\textbf{#1}}}
\newcommand{\mc}[1]{\mathcal{#1}}
\newcommand{\cH}{\mathcal{H}}
\newcommand{\cS}{\mathcal{S}}

\begin{document}

\hfill TTK-12-07

\author{Gerasimos Rigopoulos$^a$}
\author{Wessel Valkenburg$^b$}
\affiliation{$^a$
Institute for Theoretical Particle Physics and Cosmology, RWTH Aachen, D -
52056, Germany\\
$^b$
Institut f\"ur Theoretische Physik, Universit\"at Heidelberg,
Philosophenweg 16, 69120 Heidelberg, Germany}

\title{A non-linear approximation for perturbations in $\Lambda$CDM}

\begin{abstract}
We describe inhomogeneities in a $\Lambda$CDM universe with a gradient series expansion and show that it describes the gravitational evolution far into the non-linear regime and beyond the capacity of standard perturbation theory at any order. We compare the gradient expansion with exact inhomogeneous $\Lambda$LTB solutions (Lema\^itre-Tolman-Bondi metric with the inclusion of a cosmological constant) describing growing structure in a $\Lambda$CDM universe and
find that the expansion approximates the exact solution well, following the collapse of an over-density all the way into a singularity.
\end{abstract}

\maketitle

\section{Introduction}

The study of cosmological inhomogeneities through perturbation theory has been one of the cornerstones of modern cosmology \cite{Mukhanov:1990me}. Indeed, the confrontation of theory with increasingly precise observations of the CMB and the large scale clustering of galaxies through the extended use of linear perturbation theory has played an important role in the emergence of the concordance $\Lambda$CDM cosmological model and has allowed meaningful speculation about the very early universe \cite{Lyth:2009zz}. At higher order, perturbation theory is also being used to probe subdominant non-linear corrections and the associated non-Gaussianity, promising to shed additional light on the primordial universe.

Although crucially important for following the development of cosmic structures in the primordial universe, and even more recent eras on large enough scales, standard perturbation theory is naturally limited. In the matter era perturbations grow, the density contrast becomes non-linear and the inhomogeneities separate themselves completely from the background expansion; at this point standard perturbation theory breaks down completely at all orders. Some recent reformulations can somewhat extend its range of validity in the quasi-linear regime \cite{Carlson:2009it} but the most complete quantitative description of gravitational clustering is necessarily achieved by Newtonian N-body simulations. However, analytic approximations are important for obtaining insights into the process of structure formation. In particular, the Zel'dovich approximation along with a range of improvements is a prominent analytic approximation scheme and can be used to understand various features of the resulting cosmic web \cite{Sahni:1995rm}.

In this paper we use an expansion different from well known and developed cosmological perturbation theory. We approximate the metric perturbations of a $\Lambda$CDM universe by a series whose terms contain an increasing number of spatial gradients of the initial metric and not powers of the metric functions (or the density), with each term multiplied by a specified function of time. At leading (``zeroth'') order this is just the ``separate universe'' approximation. However, the inclusion of higher order terms allows for the subhorizon evolution to be traced as well and already at the first non-trivial order the approximation goes beyond standard perturbation theory and can describe the non-linear regime with gravitational collapse/void rarefaction. The inclusion of even higher orders approximates the true evolution to an increasing degree, at least if followed for a characteristic timescale. It is interesting to note that the gradient expansion as will be used below can reproduce the Zel'dovich approximation for the density contrast but also allows for relativistic corrections to be included \cite{Croudace:1993yt, Stewart:1994wq}; the latter are indeed crucial for increased accuracy during the later stages of gravitational collapse.

The gradient expansion can be used to evolve any initial configuration of perturbations, for example those described by a post-inflationary scale invariant spectrum. In this work, after presenting the general formulation, we compare the first two non-trivial terms in the expansion with exact Lema\^itre-Tolman-Bondi~\cite{1933ASSB...53...51L,Tolman:1934za,Bondi:1947av} solutions with the inclusion of a cosmological constant ($\Lambda$LTB) from Ref.~\cite{Valkenburg:2011tm} in order to assess the accuracy of the gradient series approximation. We find the approximation to provide a fairly good description, certainly much beyond the capacity of standard perturbation theory. We include terms with up to 4 spatial gradients and the inclusion of higher order terms is expected to fare even better in accurately describing the non-linear gravitational evolution.

\section{The inhomogeneous $\Lambda$CDM Universe as a series in spatial gradients}
\label{grad}
The gradient expansion as a method for obtaining solutions to the Einstein Equations dates back to \cite{Lifshitz:1963ps} and has been used as a means for approaching the initial singularity of an inhomogeneous universe\cite{Belinsky:1970ew}. It was discussed in more general terms in \cite{Tomita:1975kj, Comer:1994np} and more recently applied to inflation in \cite{Tanaka:2006zp,Tanaka:2007gh}. A leading order covariant computation can be found in \cite{Bruni:2003hm}. It as also proved useful in studying the backreaction of cosmic structure formation, see \cite{Kolb:2005da} and more recently \cite{Enqvist:2011aa}. In most of past works the gradient expansion has been applied directly on the Einstein equations. However, in \cite{Croudace:1993yt, Stewart:1994wq} a Hamilton-Jacobi (HJ) approach was used which arguably offers a less laborious way to perform the expansion. In this section we follow the HJ methods to obtain a gradient expansion for inhomogeneities in a $\Lambda$CDM universe.

The applicability of a gradient expansion does not rely on some quantity being a small perturbation around a given background but would generally require the existence of a hierarchy between spatial and temporal derivatives $\partial_t Q>\partial_i Q \sim \frac{\Delta Q}{\Delta t}>\frac{\Delta Q}{L}$
where $Q$ is either a matter or a metric quantity with $\Delta t$ and $L$ the characteristic time-scale and length-scale of the system respectively. One would generically expect the gradient expansion to be applicable on length-scales $L \gg \Delta t$ i.e. scales beyond causal contact. This has indeed been used in early universe cosmology where $\partial_t Q \sim H Q$ under the guise of the super-horizon ($\frac{1}{aH}<{\rm x}$) "separate universe" approximation which is nothing but the lowest order in a gradient expansion \cite{Rigopoulos:2003ak}. However, the gradient expansion can also be used for length scales within causal contact provided that the spatial variation of the initial data is smooth enough and the spatial gradients small. One could then still use the gradient expansion but for a limited time span: bigger spatial variations would correspond to shorter times for which the gradient series would apparently converge. We will make these statements more precise below.

\subsection{Hamilton-Jacobi for $\Lambda$CDM}

Let us start by writing an action for gravity with a cosmological constant $\Lambda$ and Cold Dark Matter (CDM)
\beq
\cS=\int d^4x\sqrt{-g}\left[\frac{1}{2\kappa}\left(^{(4)}R-2\Lambda\right)-\frac{1}{2}\rho\left(g^{\mu\nu}\partial_\mu\chi\partial_\nu\chi+1\right)\right]\,,
\eeq
where $\chi$ is a potential for the 4-velocity of CDM, $U^\mu=-g^{\mu\nu}\partial_\nu\chi$, and $\rho$ is the energy density. The latter acts as a Lagrange multiplier whose variation ensures that $U^\mu U_\mu=-1$. Variation with respect to $\chi$ gives the continuity equation $\nabla_\mu(\rho\partial^\mu\chi)=0$ while variation of the CDM part of the action with respect to $g^{\mu\nu}$ gives the usual CDM energy momentum tensor $T_{\mu\nu} = \rho U_\mu U_\nu$. Gravity is described by the standard Einstein-Hilbert term with the addition of a cosmological constant with $^{(4)}R$ the 4-dimensional Ricci Scalar, and $\kappa=8\pi G$. The ADM decomposition for the metric can now be used to develop a Hamiltonian formalism for this system. Writing the metric as
\beq
  g_{00}= -N^2+h_{ij}N^iN^j \,,\qquad
  g_{0i}= \gamma_{ij}N^j \,,\qquad
  g_{ij}= \gamma_{ij}
  \,,
 \label{ADM metric}
\eeq
and defining the canonical momenta
\bea
\pi^{ij}&\equiv&\frac{\delta \cS}{\delta\dot{\gamma}_{ij}}=\frac{\sqrt{\gamma}}{2\kappa}\frac{E^{ij}-\gamma^{ij}E}{N}\,,\\
\pi^\chi&\equiv&\frac{\delta \cS}{\delta\dot{\chi}}=\rho\sqrt{\gamma}\sqrt{1+\gamma^{ij}\partial_i\chi\partial_j\chi}\,,
\eea
with
\beq
 E_{ij} = \frac{1}{2} \dot \gamma_{ij} - \nabla_{(i}N_{j)}
\,,\qquad
 E =  h^{ij}E_{ij}
\,,
 \label{Eij-E}
\eeq
we can write the action as
\beq\label{canonical1}
\cS=\int d^4x\left(\pi^\chi\frac{\partial\chi}{\partial t}+\pi^{ij}\frac{\partial \gamma_{ij}}{\partial t}-N\mathcal{U}-N_i\mathcal{U}^i\right)\,,
\eeq
where
\beq
\mathcal{U} = \frac{2\kappa}{\sqrt{\gamma}}
                     \left(\pi_{ij}\pi^{ij}-\frac{\pi^2}{2}
                     \right) - \frac{\sqrt{\gamma}}{2\kappa}(R-2\Lambda)+\pi^\chi\sqrt{1+\gamma^{ij}\partial_i\chi\partial_j\chi}\,,
\eeq
and
\beq
\mathcal{U}_i=-2\nabla_k\pi_i^k+\pi^\chi\partial_i\chi\,.
\eeq
Choosing the velocity potential $\chi$ to define the time hypersurfaces of the system implies that $\partial_i\chi=0$ along with $\frac{\partial\chi}{\partial t}=1$ and $N=1$. The metric then takes the form
\beq
ds^2=-dt^2+\gamma_{ij}(t,\vc{x})dx^idx^j\,,
\eeq
and the spatial coordinate lines comove with the matter - there is no peculiar velocity. This choice then corresponds to the comoving synchronous gauge\footnote{There is conceptual similarity of this choice to the Lagrangian picture of cosmological perturbations \cite{Buchert:1992ya}; see also the discussion below on the relation to the Zel'dovich approximatiton.}. It also means that $\pi^\chi$ now plays the role of the Hamiltonian density:$-\pi^\chi\equiv \cH$. By imposing the energy constraint $\mathcal{U}=0$ we obtain in this gauge a hamiltonian density for the gravitational field
\beq
\cH =\frac{2\kappa}{\sqrt{\gamma}}
                     \pi_{ij}\pi^{kl}\left(\gamma_{ik}\gamma_{jl}-\frac{1}{2}\gamma_{ij}\gamma_{kl}
                     \right) - \frac{\sqrt{\gamma}}{2\kappa}(R-2\Lambda)\,,
\eeq
where R is the Ricci scalar of the 3-dimensional time hypersurfaces. The action finally takes the form
\beq\label{canonical2}
\cS=\int dt d^3x\left(\pi^{ij}\frac{\partial \gamma_{ij}}{\partial t}-\cH+ 2N_i\nabla_k\pi^{ki}\right)\,,
\eeq
and defines a constrained Hamiltonian system where the canonical momentum $\pi^{ij}$ is constrained to be covariantly conserved
\beq\label{cov-cons}
\nabla_k\pi^{ki}=0\,.
\eeq

Let us now apply the HJ approach to the Hamiltonian system (\ref{canonical2}) - see eg \cite{LL}. The action can be considered as a function of the metric and time $\cS[t,\gamma_{ij}]$. Then the momentum is simply the variation of this action with respect to the metric $\pi^{ij}={\delta \cS}/{\delta\gamma_{ij}}$, the Hamiltonian is minus the rate of change of the action with time and the HJ equation
\beq\label{HJ}
\frac{\partial \cS}{\partial t}+\int d^3x \left\{\frac{2\kappa}{\sqrt{\gamma}}
                    \frac{\delta \cS}{\delta\gamma_{ij}} \frac{\delta \cS}{\delta\gamma_{kl}} \left(\gamma_{ik}\gamma_{jl}-\frac{1}{2}\gamma_{ij}\gamma_{kl}
                     \right) - \frac{\sqrt{\gamma}}{2\kappa}(R-2\Lambda)\right\}=0\,,
\eeq
contains all the dynamical information of the system. It is a single partial differential equation for $\cS$ as a functional of $\gamma_{ij}$ and a function of $t$. Once $\cS[t,\gamma_{ij}]$ is determined the metric can be obtained from
\beq\label{dotgamma}
\frac{\partial\gamma_{ij}}{\partial t}=\frac{2}{\sqrt{\gamma}}\frac{\delta\cS}{\delta\gamma_{kl}}\left(2\gamma_{ik}\gamma_{jl}-\gamma_{ij}\gamma_{kl}\right)\,.
\eeq
Furthermore, the remaining constraint (\ref{cov-cons}) will be automatically satisfied if
\beq
\cS=\int d^3x \sqrt{\gamma}\,\mathcal{F}(t,\gamma_{ij})\,,
\eeq
where $\mathcal{F}(t,\gamma_{ij})$ is some scalar function of the metric making $\cS$ invariant under 3-D diffeomorphisms. Indeed, the variation of such a functional with respect to the metric will yield a covariantly conserved tensor which will thus satisfy (\ref{cov-cons}). Once a solution is obtained, the CDM density can be obtained from
\beq
\rho(t,\vc{x})=\frac{\rho_0}{\sqrt{\gamma(t,\vc{x})}}\,,
\eeq
where $\rho_0$ is a constant. Equations (\ref{HJ}) and (\ref{dotgamma}) are the basic equations that need to be solved in this approach.

\subsection{The gradient expansion}

The gradient expansion in this formulation begins by writing an ansatz for $\mathcal{F}$ as \cite{Croudace:1993yt,Stewart:1994wq}
\beq\label{gradient-exp}
\mathcal{F}=-2H(t)+J(t)R+L_1(t)R^2+L_2(t)R^{ij}R_{ij}+\ldots\,,
\eeq
a series in powers of the 3-D Ricci curvature (involving an increasing number in gradients of $\gamma_{ij}$). The HJ equation (\ref{HJ}) can now be solved separately for terms with a different number of gradients. This gives for the time dependent coefficients in (\ref{gradient-exp})
\bea\label{H}
\frac{dH}{dt}+\frac{3}{2}H^2-\frac{\Lambda}{2}&=&0\,,\\
\frac{dJ}{dt}+J H-\frac{1}{2}&=&0\,,\\
\frac{dL_1}{dt}-L_1 H-\frac{3}{4}J^2&=&0\,,\\
\frac{dL_2}{dt}-L_2 H+2J^2&=&0\,,
\eea
e.t.c.
The solution for $H$ then determines all other functions. For $\Lambda \neq 0$ equation (\ref{H}) gives
\beq
H=\sqrt{\frac{\Lambda}{3}}\coth\left(\frac{\sqrt{3\Lambda}}{2}\,\,t\right)\,,
\eeq
As the notation anticipated, $H$ is simply the background Hubble parameter. The other functions are then given by
\bea
J(t)&=&e^{-\int^t_{t_i} \!\!H}\int\limits^t_{t_i}\frac{1}{2}\,e^{\int^x_{t_i} \!\!H}dx=\frac{1}{\sqrt{3\Lambda}\left[\sinh{\frac{\sqrt{3\Lambda}}{2}}\,t\right]^{2/3}}\int\limits_{{\frac{\sqrt{3\Lambda}}{2}}\,t_i}^{{\frac{\sqrt{3\Lambda}}{2}}\,t}\left(\sinh u\right)^{2/3}du\,,\\
L_1(t)&=&\frac{3}{4}L(t)\,,\quad L_2(t)=-2L(t)\,,
\eea
where 
\bea
L(t)=e^{\int^t_{t_i} \!\!H}\int\limits^t_{t_i}\left[e^{-\int^x_{t_i} \!\!H}J^2\right]dx &=&{\frac{2}{\sqrt{3\Lambda}}}\left[\sinh{\frac{\sqrt{3\Lambda}}{2}}\,t\right]^{2/3}\int\limits_{{\frac{\sqrt{3\Lambda}}{2}}\,t_i}^{{\frac{\sqrt{3\Lambda}}{2}}\,t}\left(\sinh u\right)^{-2/3}J^2(u)\,du\,,\\
\eea
 $t_i$ is the initial time and $u=\frac{\sqrt{3\Lambda}}{2}\,t$.
Given the above, the metric can now be obtained by solving equation (\ref{dotgamma}). Up to terms with 4 gradients we obtain
\bea\label{dotgamma-4}
\frac{\partial\gamma_{ij}}{\partial t}&=&2H\gamma_{ij} + J(R\gamma_{ij}-4R_{ij})\nonumber\\
&&\hspace{-0.3cm}+L_1\left(3\gamma_{ij}R^2-8RR_{ij}+8R_{| ij}\right)\nonumber\\
&&\hspace{-0.3cm}+L_2\Big(3\gamma_{ij}R^{kl}R_{kl}-8R_{ik}R^k{}_j+3\gamma_{ij}R^{|k}{}_{|k}\nonumber\\
&&\hspace{-0.3cm}+4R_i{}^k{}_{|jk}+4R_j{}^k{}_{|ik}-4R_{ij}{}^{|k}{}_{|k}-4\gamma_{ij}R^{kl}{}_{|kl}
\Big)+\mathcal{O}(6)\,. \label{eq:4thorderdiffeq}
\eea

So far, the resulting Eq.~\eqref{eq:4thorderdiffeq} is exact if one includes all orders. The gradient approximation now consists of solving this equation iteratively for $\gamma_{ij}$. At zeroth order in gradients we have
\beq\label{metric-0}
\frac{\partial\gamma^{(0)}_{ij}}{\partial t}=2H\gamma^{(0)}_{ij}\,,
\eeq
which gives
\beq\label{metric-0}
\gamma_{ij}^{(0)}(t)=A^2(t)k_{ij}\,,
\eeq
where $k_{ij}$ is the initial metric at time $t=t_i$ and
\beq
A(t)=e^{\int^t_{t_i}\!H}=\left[\frac{\sinh {\frac{\sqrt{3\Lambda}}{2}}\,t}{\sinh {\frac{\sqrt{3\Lambda}}{2}}\,t_i}\right]^{2/3}\,.
\eeq
This is simply the separate universe approximation where each point evolves as a homogeneous FRW universe. Up to second order in gradients we get
\beq\label{metric-2}
\frac{\partial\gamma_{ij}^{(2)}}{\partial t}=2H\gamma^{(2)}_{ij} + J(\hat{R}k_{ij}-4\hat{R}_{ij})
\eeq
where we have substituted the $0^{\rm th}$ order result in the terms that contain two gradients: $\hat{R}_{ij}\equiv\hat{R}_{ij}(k_{lm})$ and $\hat{R}\equiv\hat{R}(k_{lm})$ are the Ricci tensor and Ricci scalar of the initial metric $k_{ij}$. The solution of (\ref{metric-2}) gives
\beq
\gamma^{(2)}_{ij}=A^2(t)k_{ij}+\lambda(t)(\hat{R}k_{ij}-4\hat{R}_{ij})\,,
\eeq
where
\beq
\lambda(t)=e^{2\int^t_{t_i} \!\!H}\int\limits^t_{t_i}\left[e^{-2\int^x_{t_i} \!H}J\right]dx = {\frac{2}{\sqrt{3\Lambda}}}\left[\sinh{\frac{\sqrt{3\Lambda}}{2}}\,t\right]^{4/3}\int\limits_{{\frac{\sqrt{3\Lambda}}{2}}\,t_i}^{{\frac{\sqrt{3\Lambda}}{2}}\,t}\left(\sinh u\right)^{-4/3}J(u)\,du\,.
\eeq
Proceeding to 4 gradients we have
\bea
\label{metric-4}
\frac{\partial\gamma_{ij}^{(4)}}{\partial t}&=&2H\gamma_{ij} + J(\hat{R}k_{ij}-4\hat{R}_{ij})\nonumber\\
&&\hspace{-0.3cm}+\,C_1\hat{R}^2k_{ij}+C_2\hat{R}^{kl}\hat{R}_{kl}k_{ij}+C_3\hat{R}\hat{R}_{ij}+C_4\hat{R}_{ik}\hat{R}^k{}_j\nonumber\\
&&\hspace{-0.3cm}+\,D_1\hat{R}^{|k}{}_{|k}k_{ij}+D_2\hat{R}_{| ij}+D_3R_{ij}{}^{|k}{}_{|k}\,,
\eea
where a vertical bar denotes covariant derivation with respect to the seed metric $k_{ij}$ and the Bianchi identity has been used to eliminate some terms. The $C$ and $D$ coefficients then read
\bea
C_1&=&8\frac{\lambda J }{A^2}-\frac{23}{4}\frac{L}{A^2}\,,\quad C_2=-12\frac{\lambda J }{A^2} + 10\frac{L}{A^2}\,,\quad C_3=-28\frac{\lambda J }{A^2}+18\frac{L}{A^2}\,,\quad C_4=48\frac{\lambda J }{A^2}-32\frac{L}{A^2}\,,\nonumber\\
D&\equiv& 2\frac{\lambda J }{A^2}-2\frac{L}{A^2}=D_1=D_2=-\frac{1}{4}D_3\,.
\eea
We obtain
\bea
\gamma_{ij}^{(4)}&=&A^2(t)k_{ij} + \lambda(t)(\hat{R}k_{ij}-4\hat{R}_{ij})\nonumber\\
&&\hspace{-0.3cm}+\,\left(A^2(t)\int\limits^t\frac{C_1}{A^2}du\right)\,\hat{R}^2k_{ij}
+\left(A^2(t)\int\limits^t\frac{C_2}{A^2}du\right)\,\hat{R}^{kl}\hat{R}_{kl}k_{ij}
+\left(A^2(t)\int\limits^t\frac{C_3}{A^2}du\right)\,\hat{R}\hat{R}_{ij}
+\left(A^2(t)\int\limits^t\frac{C_4}{A^2}du\right)\,\hat{R}_{ik}\hat{R}^k{}_j\nonumber\\
&&\hspace{-0.3cm}+\,\left(A^2(t)\int\limits^t\frac{D}{A^2}du\right)\,\left(\hat{R}^{|k}{}_{|k}k_{ij}-4\hat{R}_{ij}{}^{|k}{}_{|k}+\hat{R}_{| ij}\right)\,\,.\label{eq:4thordersolution}
\eea
One could proceed in this manner to obtain higher order terms.\footnote{A gradient expansion solution equivalent to (\ref{eq:4thordersolution}) was also obtained in \cite{Wiseman:2010sz} using a method different from the Hamilton-Jacobi approach utilized here. It would be interesting to further explore the relation with that approach. Our thanks to the referee for pointing it out.}

\subsection{Range of validity and the relation to the Zel'dovich approximation}
Before proceeding further it is worth estimating the range over which the gradient expansion can be applied. Focusing at early enough times the cosmological constant doesn't play a role and the universe evolves as if it were matter dominated. The metric in the gradient expansion then reads
\bea
\gamma_{ij}&\simeq& \left(\frac{t}{ t_{\rm i}}\right)^{4/3}k_{ij} + \frac{9}{20}\left(\frac{t}{t_i}\right)^{2}t_i^2\left[\hat{R}k_{ij} - 4\hat{R}_{ij}\right]\nonumber\\ &&+ \frac{81}{350}\left(\frac{t}{t_i}\right)^{8/3}t_i^{4}\left[\left(-4\hat{R}^{lm}\hat{R}_{lm}+\frac{5}{8}\hat{R}_{
|k}{}^{|k}+\frac{89}{32}\hat{R}^2\right)k_{ij} -10\hat{R}\hat{R}_{ij}
+17\hat{R}^l{}_i\hat{R}_{lj}-\frac{5}{2}\hat{R}_{ij|k}{}^k+\frac{5}{8}\hat{R}_{|ij}     \right]\,,
\eea
where we have ignored terms that are subdominant when $t\gg t_{\rm i}$. The 2-gradient and 4-gradient terms become of the same order of magnitude as the 0th  order one after a time $t\sim t_{\rm con}$
where
\beq\label{t-con}
t_{\rm con}\sim \mathcal{O}(\text{few})\frac{1}{t_i^2\hat{R}^{3/2}} \quad \text{or} \quad t_{\rm con}\sim\mathcal{O}(\text{few})\frac{1}{t_i^2(\nabla^2\hat{R})^{3/4}}  \,.
\eeq
The exact coefficient is of course dependent on the form of the initial 3-metric $k_{ij}$. We can therefore estimate that the expansion should be valid for $t\lesssim t_{\rm con}$. After this time the contributions to the metric of successive terms containing more gradients are no longer perturbatively ordered and the iterative solution to (\ref{dotgamma-4}) breaks down. However, note that the timescale over which the series breaks down is commensurate with the time of collapse for a region of positive spatial curvature (where $\hat{R}k_{ij} - 4\hat{R}_{ij}<0$). After that the region either forms a singularity or a caustic is created which would need extra physics for its resolution. Either way, one would not expect the current description to hold anymore and thus the limited timescale over which the expansion holds does not appear to be a serious limitation. For negative curvature regions which keep on expanding things are less clear-cut since there is no natural limit on the evolution timescale. Nevertheless, it would seem that the gradient expansion can provide a fair description for times longer than $t_{\rm con}$ even in this case - see \cite{Enqvist:2011aa}.

Let us now apply the above to our universe. Assuming the standard inflationary initial conditions, the initial seed metric takes the form
\beq
k_{ij}=\left(\frac{t_{\rm i}}{t_0}\right)^{4/3}\delta_{ij}\left(1+\frac{10}{3}\Phi(\vc{x})\right)\,,
\eeq
where $\Phi(\vc{x})$ is the primordial Newtonian potential and we have scaled the seed metric such that in the absence of perturbations the scale factor would be unity today. Note that the lowest order in the above expansion is simply the separate universe approximation where each spatial point scales as a FRW, CDM, universe with scale factor $a\propto t^{2/3}$. With $k_{ij}$ as the initial metric, we have
\beq \label{metric-1-phi}
\gamma_{ij}\simeq\left(\frac{t}{ t_0}\right)^{4/3}\left[k_{ij}+3\,\left(\frac{t}{t_0}\right)^{2/3}{t}_0^{2}\Phi_{,ij}+\left(\frac{t}{t_0}\right)^{4/3}t_0^{4}\hat{B}_{ij}\right]
+\mc{O}(6)\,,
\eeq
where
\beq\label{eq:Bdef}
\hat{B}_{ij}=\frac{9}{28}\left[19\Phi_{,il}\Phi^{,l}{}_{,j}-12\Phi_{,ij}\Phi^{,l}{
}_{,l}
+3\delta_{ij}\left((\Phi^{,l}{}_{,l})^2-\Phi_{,lm}\Phi^{,lm}\right)\right]\,.
\eeq
The timescale for which the above approximation for the metric is accurate is estimated to be
\beq
\frac{t_{\rm con}}{t_0}\simeq 3.4\times\frac{H_0^3}{\left(\nabla^2\Phi\right)^{3/2}}\,,
\eeq
in accordance to (\ref{t-con}). Equivalently, applying the Poisson equation $\nabla^2 \Phi = \frac{3}{2} a^2H^2 \Omega_m \delta_m$, we find an upper bound for the matter overdensity $\delta_m=\delta\rho_m/\rho_m$ today
\beq
 \delta_{m0} \simeq 1.5 / \Omega_m.
\eeq
We see that for describing an inhomogeneous universe which resembles ours for the whole of $t_0$, the metric can include perturbations down to about $k=0.3 h\, {\rm Mpc}^{-1}$, or $\delta_{m0}\simeq 5$. Of course, even shorter scales can be accurately described but for shorter times, comparable to the collapse times of the over-dense regions.

Given these approximations, the local density of matter can be obtained
from
\beq\label{density}
\rho(t,\vc{x})=\frac{1}{6\pi G \, t^2}\frac{1}{\sqrt{{\rm Det} \left[\delta_{ij}
+3\left(\frac{t}{t_0}\right)^{2/3}{t}_0^{2}\Phi_{,ij}+\left(\frac{t}{t_0}\right)^{4/3}t_0^{4}\hat{B}_{ij}\right]}}\,.
\eeq
Note that if this expression is expanded to linear order in $\Phi$ one recovers the result
of linear perturbation theory for the density contrast in the synchronous gauge. Dropping the $\hat{B}_{ij}$ term we have
\beq
\rho(t,\vc{x})=\frac{1}{6\pi G \, t^2}\frac{1}{{{\rm Det} \left[\delta_{ij}
+\frac{3}{2}\left(\frac{t}{t_0}\right)^{2/3}{t}_0^{2}\Phi_{,ij}\right]}}\,,
\eeq
which reproduces the well known Zel'dovich approximation. Thus, the gradient expansion in the comoving synchronous gauge can be thought of as providing a relativistic extension of the Zel'dovich approximation \cite{Croudace:1993yt,Stewart:1994wq}\footnote{Planar solutions in the presence of $\Lambda$ akin to Zel'dovich pancakes are described in \cite{Barrow:1984zz} and \cite{1989CQGra...6.1253B} for GR and Newtonian Gravity respectively}. In fact, as will also be demonstrated below, the higher order terms are crucial for increased accuracy during the more advanced stages of the gravitational evolution.

At late times $H \rightarrow \sqrt{\frac{\Lambda}{3}}$ and the universe enters a de Sitter phase in which case it is easy to see that the 4-gradient terms in the metric are exponentially suppressed, the 2-gradient terms freeze while the 0-gradient terms are exponentially enhanced. Thus, the gradient expansion becomes increasingly better with time, converging with time to the separate universe picture as would be expected, apart from patches which suffer gravitational collapse before the de Sitter phase.

\section{Comparison with $\Lambda$LTB}\label{LLTB}
\subsection{The LTB metric}
Now let us assess the quality of the Gradient Expansion by comparing it to an exact metric that approximates the formation of structure in a $\Lambda$CDM universe. The Lema\^itre-Tolman-Bondi metric is spherically symmetric, and can describe a central density perturbation embedded in an exact Friedman-Lema\^itre-Robertson-Walker (FLRW) metric by choosing an appropriate density profile. With the inclusion of dust, curvature and a cosmological constant this hence describes the rarefaction of a void as well as the initial expansion and later collapse of an over-density in a $\Lambda$CDM universe\footnote{Generalizations to spheroidal configurations can be found in \cite{1993MNRAS.262..717B}.}. The metric is given by
\begin{align}
ds^2 = dt^2 - X^2(r,t) dr^2 - Y^2(r,t) d\Omega^2,
\end{align}
with $X(r,t)\equiv \partial_r Y(r,t)/\sqrt{1+2E(r)}$. The metric is described by three free radial functions, being the curvature $E(r)$, the matter distribution $\rho(r)$ and the Big Bang time $t_{\rm BB}(r)$. Due to the gauge freedom in the radial coordinate $r$, one of the three functions can be set arbitrarily, and we choose a gauge in which $\int_0^r Y'(r,t) Y^2(r,t) \rho(r,t) dr \propto r^3$. Therefore the choice of curvature profile $E(r)$ and bang-time profile $t_{\rm BB}(r)$ defines the metric completely. Moreover, now any deviation of $t_{\rm BB}(r)$ from a constant corresponds to a purely decaying mode. Following the inflationary paradigm we know that the decaying mode must be small such that is has decayed today, and we can safely ignore it by setting $t_{\rm BB}(r)=0$ everywhere. All freedom now hence lies in $E(r)$.

The Einstein equations simplify to
\begin{align}
\left( \frac{\partial_t Y(r,t)}{Y(r,t)} \right)^2 = \left( \frac{\partial_t a(r,t)}{a(r,t)} \right)^2 = H^2(r,t)= &H_0^2\left[   \frac{1}{a^3(r,t)} +  \frac{3E(r)}{4\pi r^2 a^2(r,t)}  + \frac{\Lambda}{3H_0^2}    \right],\label{eq:LTBi}
\end{align}
where $Y(r,t)\equiv r a(r,t)$, and we see that for $E(r)\propto r^2$ we obtain exactly the FLRW solution. Moreover, if we choose $E(r)$ such that beyond a certain radius, say $r=L$, $E(r)\propto r^2$, while below that radius it has a different shape, that $E(r)$ describes an inhomogeneity embedded in FLRW. We choose,
\begin{align}
E(r) = - r^2 k W_7\left(\frac{r}{L},0\right),
\end{align}
where $k=\pm\frac{45 H_0^2}{8\pi}$ and the function $W_n(r/L,0)$ is defined in Ref.~\cite{Valkenburg:2011tm}, and interpolates from $W_n(0,0)=1$ to $W_n(1,0)=0$ while remaining $C^n$ everywhere, such that the metric is $C^{n-1}$. By choosing $n=7$ we are guaranteed to have all functions entering the problem to be at least $C^0$ everywhere. {Having $W_n(0,0)=1$ and $W_n(1,0)=0$, we have defined a curvature profile with non-zero curvature at the center of the metric at $r=0$, and zero curvature at finite radius $r=L$ such that the metric there becomes the exact spatially flat FLRW metric with dust and a cosmological constant for $r>L$.} The magnitude of $k$ is arbitrary, and we chose the value it has such that it describes a central over-density that reaches the singularity today, and an under-density that is non-linear today but does not suffer form shell crossing yet.

We use the full solution to the Einstein equation as well as $Y'(r,t)$ from Ref.~\cite{Valkenburg:2011tm} valid for any distribution of dust and curvature, until shell crossing occurs. In calculating the higher order gradient terms we encounter second spatial derivatives of the spatial Ricci scalar, which contain both $Y''(r,t)$ and $Y'''(r,t)$. Since we have exact solutions to $Y(r,t)$ and $Y'(r,t)$, we can safely evaluate the second and third radial derivative of $Y(r,t)$ numerically.

\subsection{Comparison of the gradient expansion with the exact solution}
\begin{figure}
\includegraphics[height=0.25\textheight]{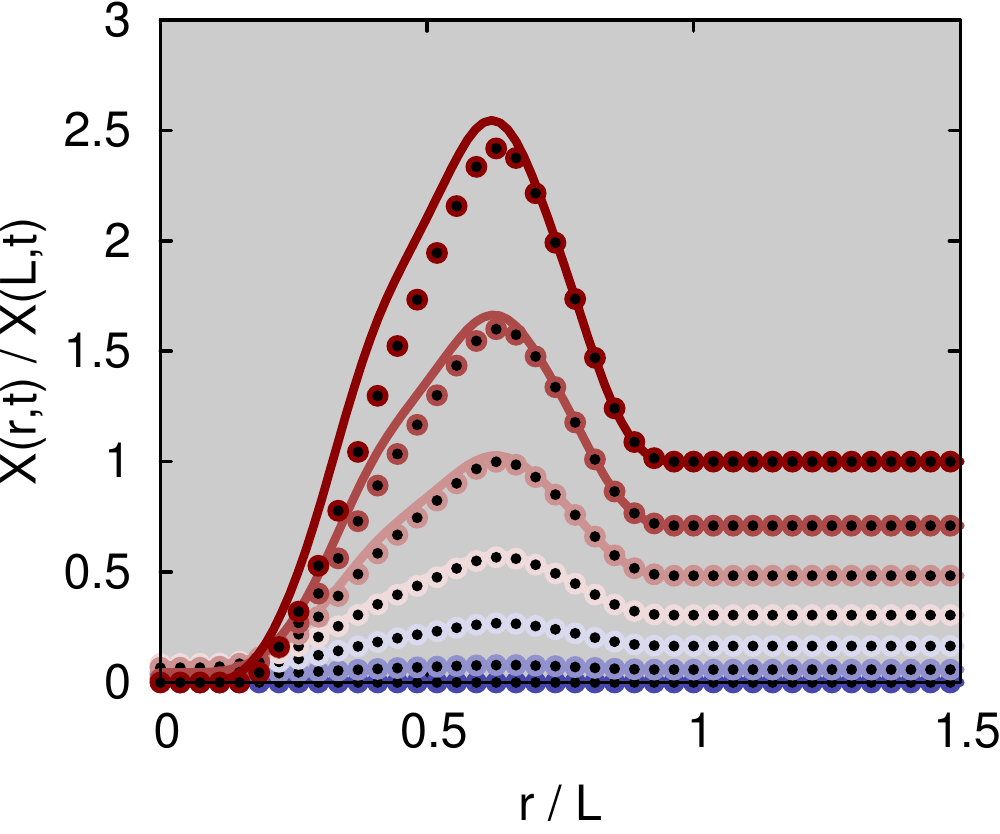}
\includegraphics[height=0.25\textheight]{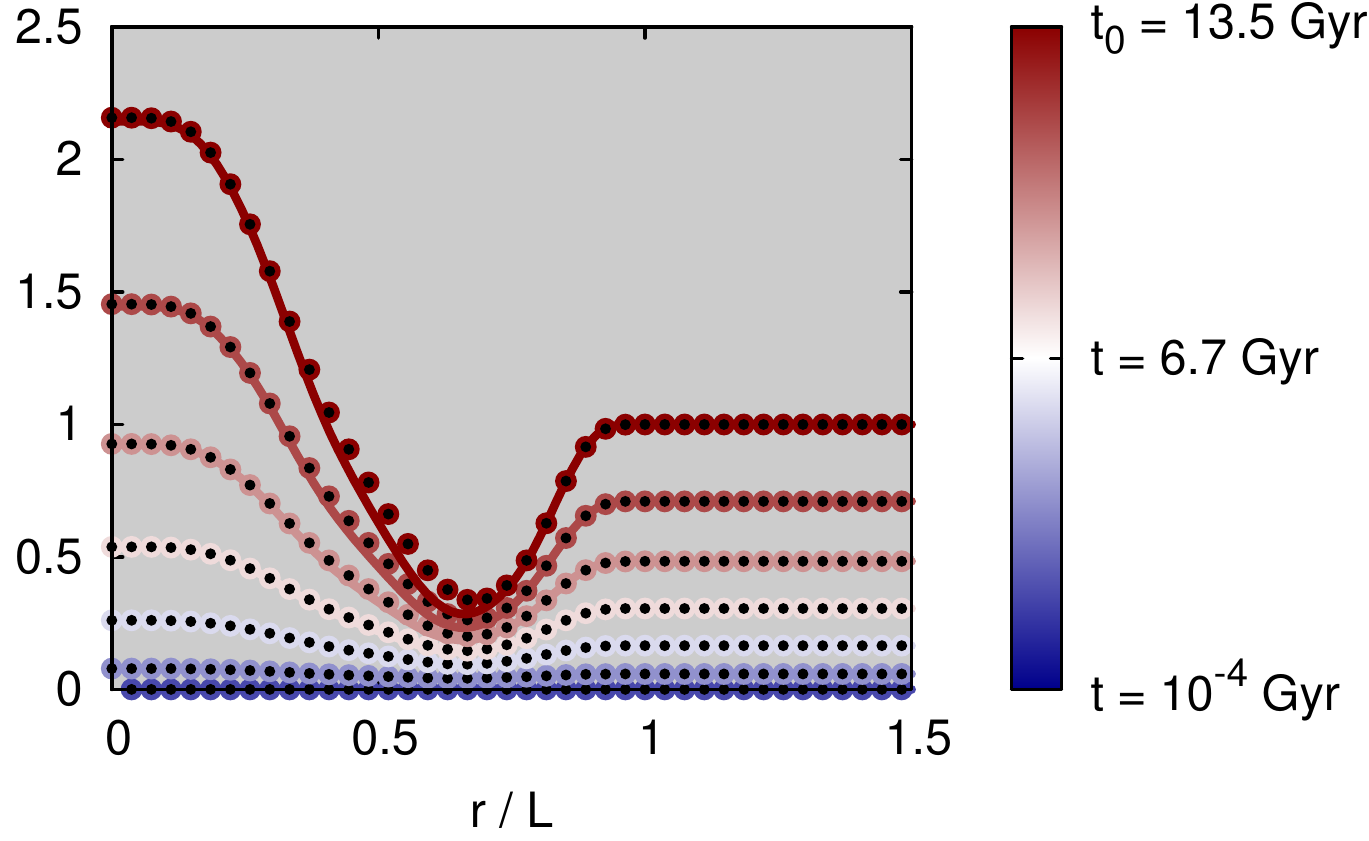}
\caption{\label{fig:fullmetricplot} The radial component of the metric $\sqrt{-g_{rr}}=X(r,t)$ as a function of radius $r$ at a number of times, where different colours correspond to different times. The solid lines are the exact solutions, the coloured dots are the gradient-expansion solutions, up to fourth order in gradients. On the left we show the solution for an over-density, compensated by an under-dense shell and matching to FLRW at exactly $r=L$. The center of this particular configuration reaches the singularity at $t=13.5$ Gyr after the initial singularity. On the right we show an under-density with the same curvature as the over-density, but opposite sign. By eye, both scenarios are very well approximated by the gradient expansion. See Figures~\ref{fig:centralcollapse}~and~\ref{fig:goodnessoffit} for quantification of this statement.
}
\end{figure}

We calculate the initial conditions to Eq.~\eqref{eq:4thorderdiffeq} from the exact metric satisfying~\eqref{eq:LTBi} at some early initial time and  numerically integrate the set of six coupled differential equations that describe $A(t)$, $J(t)$, $L(t)$, $\lambda(t)$, $\int \lambda(t) J(t) / A^2(t)\,dt$ and $\int L(t) / A^2(t)\,dt$ at any later time. We then compare the resulting solution~\eqref{eq:4thordersolution} to the exact solution from~\eqref{eq:LTBi}.

In Figure~\ref{fig:fullmetricplot} we show the results for an over-density with $k=\frac{45 H_0^2}{8\pi}$ (left) and an under-density with $k=-\frac{45 H_0^2}{8\pi}$ (right). We plot $\sqrt{-g_{rr}}=X(r,t)$ as a function of radius $r$ for different times, where time is encoded by the colouring of the lines and dots. Solid lines correspond to the exact solution, while dots present the expansion up to fourth order in gradients. The expansion follows the features in the metric very well, in both the compensating shells and the central over- and under-densities. Obviously, for $r>L$, where the metric is exactly FLRW, the gradient expansion corresponds to the exact solution with $g_{rr}\propto a^2(t)$.

\begin{figure}
\includegraphics[scale=0.4]{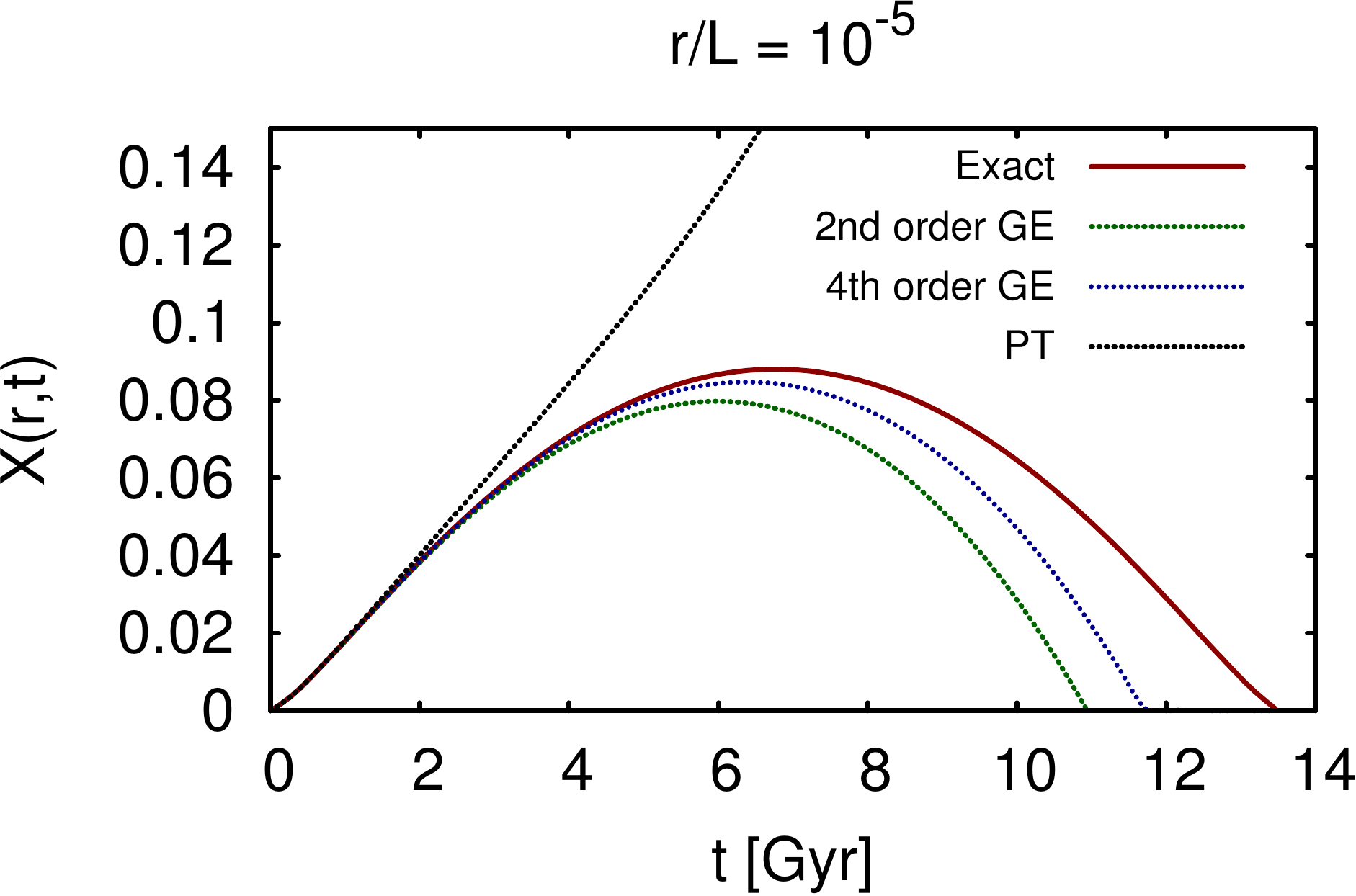}
\caption{\label{fig:centralcollapse} The radial component of the metric $\sqrt{-g_{rr}}=X(r,t)$ at $r=10^{-5}L$ (practically at the centre) as a function of time $t$, comparing the exact solution (solid red), linear perturbation theory (dotted black, uppermost line), 2nd order gradient approximation (dotted green, innermost curve) and 4th order gradient approximation (dotted blue). While linear perturbation theory clearly does not show any form of collapse, both lowest (2nd) and next order (4th) in the gradient approximation show the collapse. The inclusion of 4th order gradient terms approximates the exact solution better than the truncation to 2nd order.}
\end{figure}

To clearly show how well the gradient expansion follows the gravitational collapse of an object, we show in Figure~\ref{fig:centralcollapse} the radial metric component $\sqrt{-g_{rr}}=X(r,t)$ very close to the centre. Exactly at the centre we have $X(r,t)\equiv0 \forall t$, since it corresponds to $r \,a(t)$ in FLRW, which is why we plot $X(r,t)$ just off-centre, at $r=10^{-5}L$. For illustration we show the solution at first order in linear perturbation theory, which shows no sign of collapse. The fourth order gradient expansion follows the exact solution closer than the second order expansion, but both anyway show the collapse to a central singularity.

\begin{figure}
\includegraphics[scale=0.4]{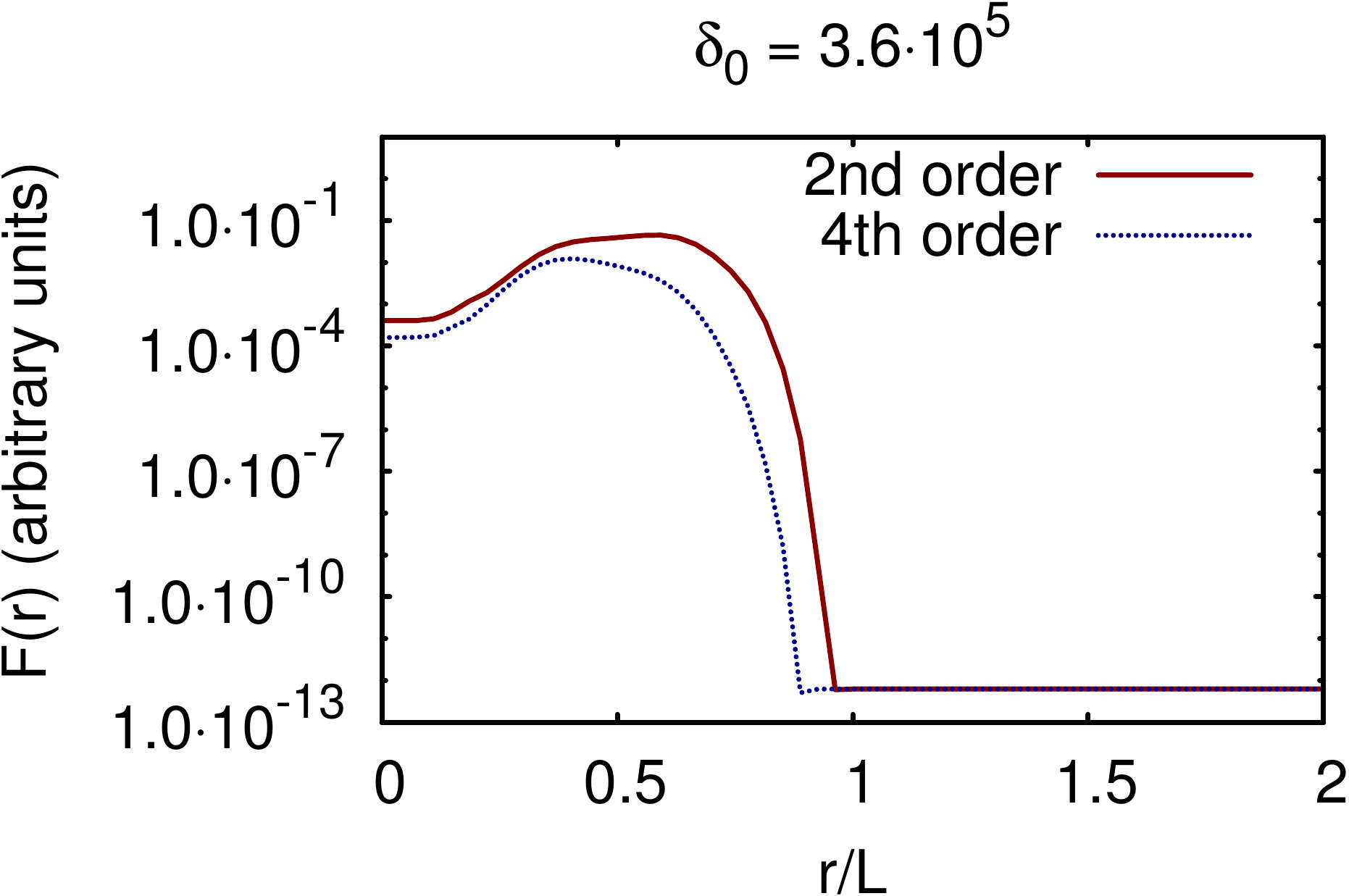}
\includegraphics[scale=0.4]{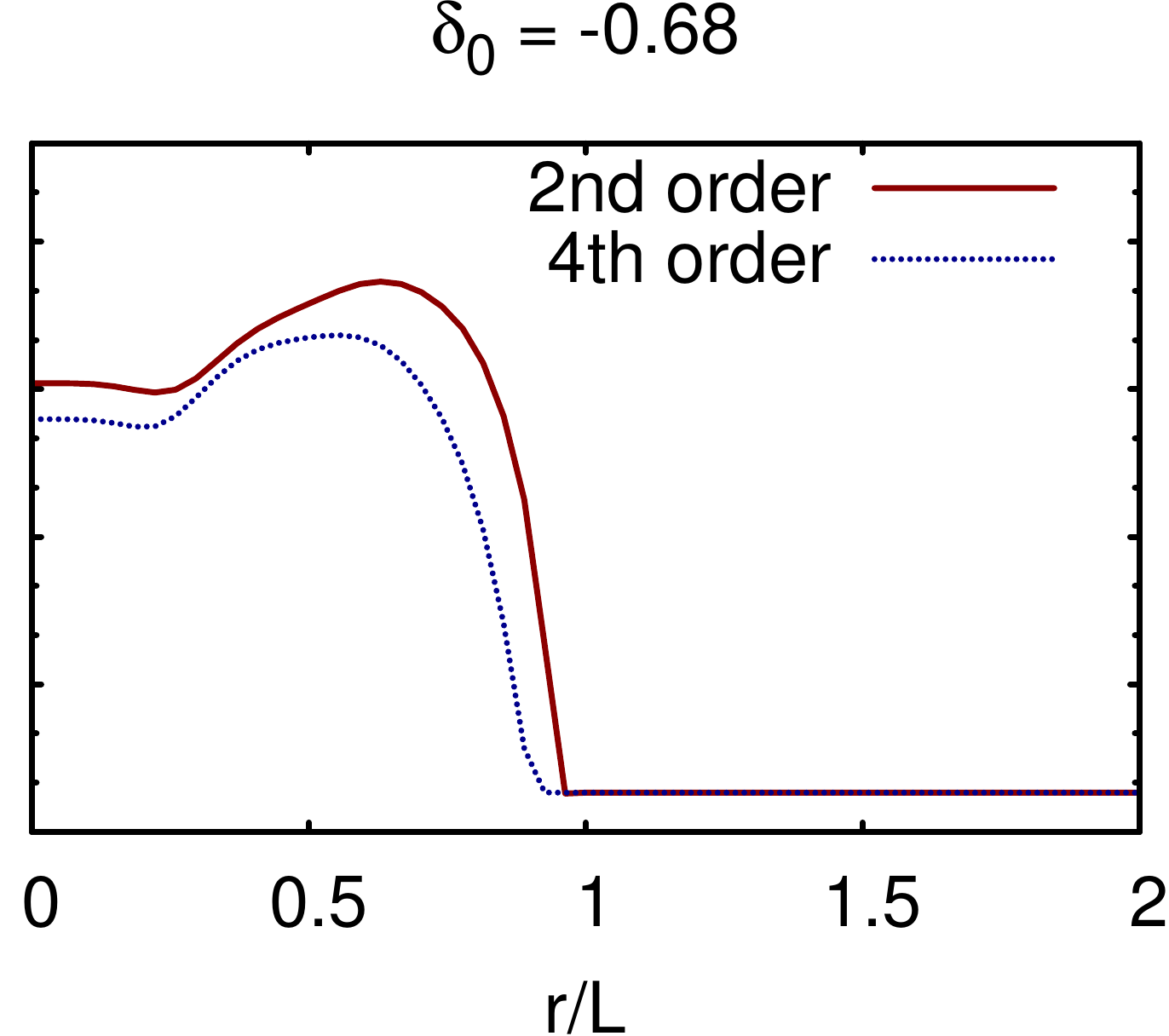}
\caption{\label{fig:goodnessoffit} The goodness of approximation $F(r)$, defined in Eq.~\eqref{eq:goodness}, for both the over- and under-density. The expansion up to 2nd order in gradients it plotted in solid red, up to 4th order in gradients in dotted blue. The units are irrelevant. What matters is the improvements of the 4th order over the 2nd order solution. See the text for an explanation of the shape of these curves.
}
\end{figure}

In order to quantify the improvement of the expansion up to fourth order in gradients over the expansion up to second order in gradients, we calculate the square of the difference between the approximation and the exact solution:
\begin{align}
    f(r,t)\equiv& \left[ X_{\rm gradient}(r,t) - X_{\rm exact}(r,t)\right]^2,\nonumber\\
    F(r)\equiv &\int_0^{t_0} f(r,t) \,dt.\label{eq:goodness}
\end{align}
One could also integrate $f(r,t)$ over time, or both time and radius, but we focus on the time integrated goodness of fit at different radii. This way we can see in Figure~\ref{fig:goodnessoffit} that the expansion up to fourth order in gradients everywhere outperforms the expansion up to second order reaching a time integrated error $\sqrt{F}$ of less than $10\%$ for all radii. In both configurations there is a point at $r\simeq 0.3 L$ where the first gradient term $\hat{R}k_{ij} - 4\hat{R}_{ij}=0$  which signifies the transition between the central region and the compensating shells. At this point the leading order correction is actually the 4-gradient term and thus offers the smallest improvement in approximation; the point behaves closest to a ``separate universe''. A better approximation at this point would presumably require the inclusion of higher order terms. The overall approximation is better in the central region where the perturbation profile is relatively flat. In the compensating shells, $r\gtrsim 0.3 R$ the value of $\sqrt{F}$ is higher but the inclusion of the 4-gradient term offers significant improvement. Eventually the approximation converges the the flat $FLRW$ metric as $r\rightarrow L$.
The reason why $F(r)$ does not go to zero for $r>L$, is simply that we are comparing a numerical integration to an exact solution, and the difference between the solutions agrees within the desired numerical accuracy in the integration process, which is an arbitrary choice.

\section{Discussion}
\label{Discussion}
The gradient expansion is an approximation scheme of a different nature from standard cosmological perturbation theory. Instead of assuming perturbations of small amplitude around a background, it builds a solution from an inhomogeneous initial seed metric $k_{ij}$ by including terms with an increasing number of gradients of $k_{ij}$ through combinations of the 3-Ricci tensor $\hat{R}_{ij}[k_{ij}]$ and its derivatives: $\gamma_{ij}\sim\sum\limits_n F_n(t)(\hat{R}^{n}_{ij}+\sum\limits_r\hat{\nabla}^{2r}{\hat{R}}^{n-r})$, where a hat indicates that quantities are computed from $k_{ij}$. The time evolution is then encapsulated in the functions $F_n(t)$. The approximation is expected to break down when $F_1(t)\hat{R}\sim 1$ but this roughly coincides with the time of collapse of over-dense regions. The ability of describing collapse and consequently the evolution in the regime when $\frac{|\delta\rho|}{\rho}>1$ clearly goes beyond the capability of conventional perturbation theory at any order. The gradient series also contains the Zel'dovich approximation in its lowest order while higher orders provide relativistic corrections which are in fact required for greater accuracy.

In this paper the gradient expansion was formulated for the case of a $\Lambda$CDM universe and offers a way to approximate structure formation in the non-linear regime. Furthermore, in order to gauge the accuracy of the approximation, we have used it on a spherically symmetric distribution of matter with a non-trivial radial profile for which exact solutions are known in terms of the LTB metric. We found the approximation to work rather well. The first non-trivial terms containing 2 gradients describe the situation qualitatively well by showing collapse to a singularity in a manner comparable to the Zel'dovich approximation. The inclusion of 4 gradients provides quantitative improvement by approaching the exact solution even better and it is expected that the inclusion of higher order terms would provide even more accuracy, an expectation that would be interesting to confirm in future work. The expansion could also be checked for less symmetric situations, perhaps by comparing with other exact solutions or numerical simulations. Even if at the end of the day one has to resort to simulations for the most accurate description of the non-linear dynamics, it would pay to have analytic approximations that capture the essential features of non-linear evolution in a relatively simple way and with some level of realism as they would be applicable to a variety of problems. For this reason gradient-expansion techniques might be be useful additions to the toolbox of cosmologists as they strive towards a more accurate description of the universe.

\section{Acknowledgements}
GR is supported by the Gottfried Wilhelm Leibniz programme of the Deutsche Forschungsgemeinschaft. WV acknowledges funding from DFG through the project TRR33 The Dark Universe.

\bibliographystyle{redpants}
\bibliography{Grad-Appr-Jrnl2}

\end{document}